\documentclass[prd,aps,superscriptaddress,10pt,nobibnotes,notitlepage,twocolumn,longbibliography,nofootinbib,amsmath,amssymb]{revtex4-1}
\usepackage{url}
\usepackage{hyperref}
\hypersetup{
    colorlinks=true,
    linkcolor=blue,
    filecolor=magenta,      
    urlcolor=blue,
    pdftitle={Overleaf Example},
    pdfpagemode=FullScreen,
    }
\newcommand{\EM}[1]{{\textcolor{blue}{[EM: #1]}}}

\usepackage{hhline}
\usepackage{enumitem}
\usepackage{subfigure}
\usepackage{graphicx}
\usepackage{dcolumn}
\usepackage[no-test-for-array]{nicematrix}
\usepackage{slashed}
\usepackage{bm}

\def\beq{\begin{equation}}
\def\eeq{\end{equation}}
\def\be{\begin{equation}}
\def\ee{\end{equation}}
\def\bea{\begin{eqnarray}}
\def\eea{\end{eqnarray}}

\def\te{e}

\begin{document}
\title{The \texorpdfstring{$\pi$}--axion and \texorpdfstring{$\pi$}--axiverse of dark QCD}
\medskip\
\author{Stephon Alexander}%
\email[Email: ]{ stephon\_alexander$@$brown.edu}
\affiliation{Department of Physics,
Brown University, Providence, RI 02912, USA}
\author{Humberto Gilmer} 
\email[Email: ]{ humberto\_gilmer$@$brown.edu}
\affiliation{Department of Physics,
Brown University, Providence, RI 02912, USA}
\author{Tucker Manton} 
\email[Email: ]{ tucker\_manton$@$brown.edu}
\affiliation{Department of Physics,
Brown University, Providence, RI 02912, USA}
\author{Evan McDonough} 
\email[Email: ]{ e.mcdonough$@$uwinnipeg.ca}
\affiliation{Department of Physics, University of Winnipeg, Winnipeg MB, R3B 2E9, Canada}

\begin{abstract}
Axions and axion-like particles (ALPs) are a prominent dark matter candidate, drawing motivation in part from the axiverse of string theory.  Axion-like particles can also arise as composite degrees of freedom of a dark sector, for example, as dark pions in dark Quantum Chromo-Dynamics. In a dark Standard Model (SM) wherein all 6 quark flavors are light while the photon is massive, one finds a rich low-energy spectrum of stable and ultralight particles, in the form of neutral and charged dark scalars, and complex neutral scalars analogous to the SM kaon, with mass splittings determined by the mass and charge of the dark quarks.  The model finds a natural portal to the visible sector via kinetic coupling of the dark and visible photons, and consequent millicharges for dark matter. The dark matter can be a mixture of all these ultralight bosonic degrees of freedom, and exhibit both parity-even and parity-odd interactions, making the theory testable at a wide variety of experiments.  In context of dark QCD with $N_f$ flavors of light quarks, this scenario predicts $N_f^2-1$ ultralight axion-like particles — effectively an axiverse from dark QCD. This `$\pi$-axiverse' is consistent with but makes no recourse to string theory, and is complementary to the conventional string theory axiverse.
\end{abstract}

\maketitle

\tableofcontents

\section{ Introduction}

Axions and axion-like particles have entered center stage as a candidate for physics beyond the Standard Model and as dark matter. Originally motivated by the strong {\sf CP} problem of the Standard Model  \cite{Peccei:1977hh,Wilczek:1977pj,Weinberg:1977ma},  axions are also motivated by string theory, wherein vacua of string theory are conjectured to have tens to hundreds of axion-like particles \cite{Svrcek:2006yi,Arvanitaki:2009fg,Cicoli:2012sz}. Many experiments are actively searching for axion dark matter, across the broad mass range $10^{-21}$ eV to $1$ eV (see Ref.~\cite{Marsh:2015xka} for a review of models and constraints).

In this work we consider an alternate path to axion-like particles, using the experimentally tested physics of the Standard Model. We consider a dark copy of the Standard Model wherein the dark quark masses are much lower than in the Standard Model, while the dark QCD scale is much higher. As proposed in Ref.~\cite{Maleknejad:2022gyf}, axion-like particles emerge as composite degrees of freedom, analogous to pions, in the confining phase of the dark QCD theory. The possibility that ultralight dark matter could be a composite degree of freedom has been studied in only a small number of past works, see Refs.~\cite{Alexander:2018fjp,Alexander:2020wpm,Maleknejad:2022gyf}. Analogous to the global $U(1)$ shift symmetry that protects the mass of an axion, the dark pion masses are protected by a global $SU(N_f)\times SU(N_f)$ symmetry where $N_f$ is the number of flavors of quark in the dark QCD theory.
The dark pions are cosmologically stable and can easily have masses conventionally associated with axion dark matter. 
This scenario differs from the `dynamical axion'  models \cite{Kim:1984pt,Choi:1985cb} (see also \cite{Hook:2014cda,Gherghetta:2016fhp,Croon:2019iuh,Gaillard:2018xgk}), which use confinement of a dark $SU(N)$ to solve the strong {\sf CP} problem of the Standard Model, in that it does not predict new colored particles or heavy axions, but instead utilizes ultralight dark quarks which confine into ultralight dark pions.

We will refer to dark pions that fall in the axion dark matter mass range as {\it $\pi$-axions}. In the simple setup of a dark copy of the Standard Model QCD wherein all six flavors of quark are nearly-massless,  one finds 35 $\pi$-axions. For $N_f $ flavors one finds $N_f ^2 -1$  $\pi$-axions; for $N_f\gg1$, this amounts to a $\pi$-axi{\it verse}, analogous to the string axiverse  \cite{Svrcek:2006yi,Arvanitaki:2009fg,Cicoli:2012sz}. This builds on earlier work \cite{Maleknejad:2022gyf} in the context of $N_f=2$, where dark pions were first proposed as an axion-like dark matter candidate.

This model finds a natural portal to the Standard Model in the form a kinetic coupling of the visible and dark sector photons, which generates a visible electric charge (``millicharge'') for the dark Standard Model fields including the dark quarks. This generates interactions between the charged $\pi$-axions and SM photon, and, upon integrating out the dark $W$ and $Z$ bosons, new interactions with photons for both the charged and complex neutral $\pi$-axions. These interactions, including both parity-odd and parity-even, make the model amenable to a wide array of experimental probes.

There are many possibilities for $\pi$-axion dark matter. Depending on the quark masses, the millicharge parameter $\varepsilon$, and the dark photon mass $m_{\gamma'}$, the dark matter can be a combination of all 35 light degrees of freedom (real neutral, complex neutral, and charged), or solely the neutral states, or a subset thereof. Remarkably, the neutral $\pi$-axions are a stable cold dark matter candidate even in the limit of unit charge $\varepsilon\rightarrow 1$, due to the combination of a small $\pi$-axion mass and large decay constant. The mass splitting between the neutral $\pi$-axions is set by the masses of the quarks, and even in case of completely degenerate quark masses, one finds a mass splitting with two real neutral $\pi$-axions with mass $m_{\pi}= \sqrt{m_q F_\pi}$, two with mass $\sqrt{2}m_\pi$, a fourth with $\sqrt{3}m_{\pi}$, and a fifth with mass $\sqrt{6}m_{\pi}$.

The $\pi$-axion dark matter scenario is easily distinguished from conventional axion scenarios: By the dense mass spectrum that follows from the approximate $SU(N_f)\times SU(N_f)$ flavor symmetry of the underlying dark QCD theory; by the charged $\pi$-axions and neutral complex $\pi$-axions that accompany the neutral real pseudoscalar $\pi$-axions; and by the combination or parity-odd and parity-even couplings allowed by the symmetries of the model. Parity-odd couplings, accessible experimentally at experiments such as ADMX, arise from real pseudoscalar $\pi$-axions, while parity-even couplings, accessible at experiments searching for oscillations of $\alpha_e$, arise from the complex neutral and charged $\pi$-axions.

The $\pi$-axiverse offers two key benefits: (1) it utilizes physics that has been observed in the lab, and (2) it is calculable and therefore predictive: while QCD physics calculations are challenging and one often must resort to the lattice, this challenge pales in comparison to the complications of string constructions, such as competing instanton effects. Finally, we note that the $\pi$-axiverse is compatible with string theory, and indeed relatively easy to engineer, due to the similarity to the Standard Model and the decades of progress in string phenomenology.

The outline of this paper is as follows: In Sec.~\ref{sec:dSM} we outline a dark Standard Model which provides an implementation of the $\pi$-axion scenario. In Sec.~\ref{sec:dQCD} we develop the details of the $\pi$-axions, including the mass spectrum and quark content, and in Sec.~\ref{sec:interactions} we develop the interactions with Standard Model photons. In Sec.~\ref{sec:cosmo} we develop the cosmology of $\pi$-axion dark matter, in Sec.~\ref{sec:detection} focus on the observable signals through both the parity-odd and parity-even portals. We conclude in Sec.~\ref{sec:discussion} with a discussion of directions for future work.

\section{ Dark Standard Model}
\label{sec:dSM}

We consider a dark copy of the Standard Model of particle physics, including the gauge group, field content, and interactions, but with parameters of the model that are independent from those in the visible sector. We couple the dark and visible sectors via an electric millicharge of the dark fermions, arising from a kinetic mixing of the dark and visible photons.

The characteristic feature of this dark Standard Model (dSM) will be two energy scales: A UV scale to which we anchor both the Higgs vacuum expectation value (VEV) $v$ and dark QCD scale $\Lambda_{\rm dQCD}$, 
\begin{equation}
    v , \Lambda_{\rm d QCD}  \gtrsim 10^{11} {\rm GeV} ,
\end{equation}
corresponding to the energy scale conventionally associated with an axion decay constant (see e.g. \cite{Marsh:2015xka}),  and an IR scale to which we anchor the dark quark masses $m_q$,
\begin{equation}
     m_{q} \ll {\rm eV} ,
\end{equation}
which can be realized via small quark Yukawa couplings, $y_q = m_q/v \ll 1$.

Meanwhile, we assume that the charged leptons (electron, muon, tau) of the dSM have masses at the UV scale, $m\sim v, \Lambda_{\rm dQCD}$. This choice is made for simplicity in our analysis of the cosmology of this model. This implies a hierarchy between the lepton Yukawa couplings and the quark Yukawa couplings. 
The smallness of the quark Yukawa couplings is technically natural \cite{tHooft:1979rat}, on the basis of the enhanced (flavor) symmetry in the massless quark limit, which guarantees that quantum corrections to the Yukawa couplings are proportional to the couplings themselves.

Altering the Higgs VEV raises possible concern about the stability of the Higgs potential. The Higgs VEV $v$ is related to the Higgs mass $m_H$ and quartic self-coupling $\lambda$ through the relation
$v \propto m_H /\sqrt{\lambda} $. 
Anchoring $v\sim10^{11}\text{ GeV}$ requires a combination of making $m$ large or $\lambda$ small. Naturalness considerations suggest that the former case is favored, as the large lepton masses would destabilize a small dark Higgs mass as in the visible hierarchy problem. In addition, the visible Higgs exists at a point of metastable equilibrium due to its mass and the Yukawa coupling to the top quark. In contrast, the dark SM as proposed here, with $m_{H}\gtrsim v$ and $\lambda<1$, sits at a point of {\it absolute} stability, due to the large mass and small self-coupling.

The dark $W$ and $Z$ bosons are anchored to the Higgs VEV, by the standard relations
\begin{equation}
    m_{W} \sim \frac{1}{2} g v \,\,  , \,\, m_{Z} \sim \frac{1}{2}\sqrt{g^2 + g'^2}v, 
\end{equation}
where $g$ and $g'$ are the dark weak and dark hypercharge coupling constants, respectively. The freedom to set $g$ and $g'$ provides some flexibility in adjusting the dark $W$ and $Z$ boson masses, which could in principle be much much less than the Higgs VEV $v$. The dark neutrinos, while possibly light, are effectively sequestered by the heavy dark electroweak sector, and will play no role in the model as studied here.

The dark photon can be either massive or massless in our model, and is kinetically coupled to the visible photon, generating millicharges for the dark quarks and leptons (see e.g.~\cite{Cline:2021itd} for a review). The dark photon mass could arise by a variety of means; e.g., through a St\"{u}ckelberg mechanism, by an additional dark Higgs field, or by modifying the Higgs sector along the lines of the Georgi-Glashow model, employing a Higgs triplet. We take the dark photon mass $m_{\gamma'}$ and kinetic coupling $\varepsilon$ to be free parameters of the model, subject to observational constraints on dark radiation (generated by annihilations of charged $\pi$-axions) and millicharged particles. For much of this work, we will focus on the case that $m_{\gamma'}\gtrsim \Lambda_{\rm dQCD}$, in which case the primary role of the dark photon is to generate the millicharges of the dark quarks.

Finally, we come to the QCD sector of the dark Standard Model, which will be the main focus of this work. Analogous to the visible sector, we group the six dark quarks $\{u,d,s,c,b,t\}$ into three generations, characterized by mass scale $m_{\mathrm{I}}, m_{\mathrm{II}}, m_{\mathrm{III}}$, and assume $m_{\mathrm{I}}\lesssim m_{\mathrm{II}}\lesssim m_{\mathrm{III}}$ without loss of generality. We write the individual dark quark masses as 
\begin{equation}
    \begin{split}
        m_u=c_1m_{\mathrm{I}} \ \ \ & \ \ \ m_d=c_2m_{\mathrm{I}}, \\
        m_s=c_3m_{\mathrm{II}}, \ \ \ & \ \ \ m_c=c_4m_{\mathrm{II}}, \\
        m_b=c_5m_{\mathrm{III}}, \ \ \ & \ \ \ m_t=c_6m_{\mathrm{III}}, \\
    \end{split}
\end{equation}
where the parameters $c_{1},\dots,c_6$ are $\mathcal{O}(1)$ numbers. We give the dark quarks the same charge assignments as their Standard Model analogs.

Some key features of the dark Standard Model considered here in relation to the dark quark sector are:
\begin{itemize}
\item A portal to the visible SM is provided via the dark photon kinetic mixing with the visible sector photon with coupling $\varepsilon$; each dark quark gains a visible Standard Model (``milli'') electric charge $\propto \varepsilon e$.
\item At low energies, the dark electroweak sector leaves traces of its full UV description in the form of otherwise-forbidden dark quark interactions, e.g. flavor-changing interactions.
\item While we focus on the regime of pre-inflationary chiral symmetry breaking, modelling our analysis on pre-inflationary Peccei-Quinn symmetry breaking axion dark matter, one could relax this assumption,
opening up new possibilities related to both the dark QCD phase transition and the dark electroweak phase transition, such as gravitational waves \cite{Tsumura:2017knk} and baryogenesis (see e.g. Refs.~\cite{Hall:2019ank,Fujikura:2021abj}).
\end{itemize}
In what follows we predominantly focus on the quark sector of the theory, including the above features, returning to explicit features of the dark Standard Model only when necessary to assess self-consistency.

\section{ The \texorpdfstring{$\pi$}--axiverse of Dark QCD }
\label{sec:dQCD}

We now focus our attention on the dark quarks and dark gluons,  amounting to a dark version of Quantum Chromo-Dynamics (QCD). The Lagrangian, for arbitrary number of gluon colors $N_c$ and quark flavors $N_f$, is given by
\begin{equation}
\label{eq:dQCD}
    \mathcal{L}_{\mathrm{dQCD}}=-\frac{1}{4}G_{\mu\nu}^aG^{\mu\nu}_a+\sum_{n=1}^{N_f}\bar{q}_{n}(i\slashed{D}-m)q_{n},
\end{equation}
where $G_{\mu\nu}^a$ is the gluon kinetic term for $SU(N_c)$ ($N_c$ is the number of colors), $\slashed{D}\equiv \gamma^\mu D_{\mu}$ is the gauge covariant derivative, 
and $m$ is the quark mass matrix generated by the Yukawa couplings, which is chosen to be diagonal. In the massless limit, the left- and right-handed quarks can be subjected to separate chiral flavor rotations, amounting to a global symmetry group 
\begin{eqnarray}
&&U(N_f)_{\rm L}\times  U(N_f)_{\rm R} = \\
&& \,\,\,\, \quad\quad SU(N_f)_{\rm L}\times SU(N_f)_{\rm R}\times U(1)_V\times U(1)_A . \nonumber
\end{eqnarray}
At low energies,  below the confinement scale $\Lambda_{\rm dQCD}$, the theory develops a $q \bar{q}$ condensate, leading to spontaneous breaking of the chiral flavor symmetry to its diagonal subgroup,
\begin{equation}
    SU(N_f)_{\text{L}}\times SU(N_f)_{\text{R}}\rightarrow SU(N_f)_V.
\end{equation}
The Goldstone bosons of chiral symmetry breaking are known as the pions, numbering in $N_f^2 -1$ of them\footnote{The $q\bar{q}$ condensate also breaks $U(1)_A$, but this symmetry is anomalous and so the corresponding pseudo-Nambu-Goldstone boson (corresponding to the $\eta'$ of the visible SM) derives its mass separately from that of the other mesons generated by this mechanism}. The quark masses in Eq.~\eqref{eq:dQCD}, generated by the Yukawa couplings of the dSM, explicitly break chiral symmetry, rendering the symmetry only approximate, and in turn leading to small masses for the pions. In this work, we refer to ultra-light pions, namely in the mass window of axion dark matter, as $\pi$-{\it axions}.

At leading order, the $\pi$-axions may be described by a $\sigma$-model, (see \cite{Kaplan:2005es,Scherer:2005ri})
\begin{equation}\label{sigmamodel}
    \mathcal{L}_{\text{eff}}=\frac{F_\pi^2}{4}\text{Tr}[(D_\mu U)(D^\mu U)^\dagger]+\frac{\langle q\bar{q}\rangle}{2}\text{Tr}[MU+M^\dagger U^\dagger],
\end{equation}
where $F_\pi$ is the $\pi$-axion decay constant, $M$ is the mass matrix, $U={\rm exp}(\frac{2\pi i}{F_\pi}\pi^a T_a)$, with $T_a$ the generators for $SU(N_f)$, $\pi^a$ the $N_f^2-1$ dimensional  vector of $\pi$-axions, and $\langle q\bar{q}\rangle$ is the condensate parameter. 

While $F_{\pi}$, $\langle q \bar{q} \rangle$, and $\Lambda_{\rm d QCD}$, all arise from the physics of confinement, their precise relation is renormalization scheme dependent.  When performing numerical estimates, we will approximate the decay constant $F_{\pi}$ by the dQCD scale
\begin{equation}
    F_{\pi} \sim \Lambda_{\rm d QCD},
\end{equation}
though we note this relation is 
expected to hold only up to ${\cal O}(1)$ numerical factors, that can be precisely computed on the lattice.

The $\pi$-axion masses, on the other hand, depend on both the quark mass scale and dark QCD scale, according to the Gell-Mann--Oakes--Renner (GMOR) relation \cite{Gell-Mann:1968hlm},
\begin{equation}\label{GMOR}
    m_\pi^2\simeq\frac{\langle q\bar{q}\rangle}{F_\pi^2}\sum_{i} m_{q_i}.
\end{equation}
where $m_{q_i}$ are the quark masses that constitute the $\pi$-axion in question and $\langle q \bar{q}\rangle \sim \Lambda_{\rm dQCD}^3$ is the quark-antiquark condensate. This expression is less accurate when a given $\pi$-axion quark composition is the sum of more than two flavors, e.g., in the case of the SM $\eta$ particle, whose flavor eigenstate involves the up, down, and strange quarks. However, even in this case, Eq.~\eqref{GMOR}  matches the observed mass up to a discrepancy of $4.6\%$, which is more than sufficient for our current purposes.

For the charged $\pi$-axions, the mass formula gets corrected due to loops of photons \cite{Kaplan:2005es}, both visible and dark. The charged $\pi$-axion mass can then be written as 
\begin{equation}
\label{eq:mpm}
    m_{\pi_i^{\pm}}^2 \simeq 
    \begin{cases}
    m_{\pi_i ^0}^2 + 2\xi_i {e'}^2 F_{\pi}^2 \,\,\, ,\,\, m_{\gamma'} < F_{\pi} \\
m_{\pi_i ^0}^2 + 2\xi_i e^2\varepsilon^2 F_{\pi}^2 \,\,\, ,\,\, m_{\gamma'}> F_{\pi} ,
\end{cases}
\end{equation}
where the two cases correspond to the dark photon dominating the loop correction vs. the dark photon being integrated out, and the loop being dominated by a visible sector photon. Here $e = \sqrt{4 \pi \alpha_e}$ is the dimensionless electric charge parameter, $e'=\sqrt{4\pi\alpha_{e'}}$ is the dimensionless dark EM coupling (where $\alpha_{e'}$ need not be $\sim\frac{1}{137}$), and $\xi_i$ is an $O(1)$ parameter unique to each $\pi$-axion, which can be positive or negative. For the majority of this work, we assume that $m_{\gamma'}>F_\pi$ in order to consider the various millicharge scenarios. We will comment on the $m_{\gamma'}<F_\pi$ case when needed, and in the discussion, as it pertains to the muon $g-2$ anomaly.

We henceforth focus on the case $N_c=3$ and $N_f=6$ in accordance with the dark Standard Model of the previous section. Using the $SU(6)$ generators from \cite{Hartanto:2005jr}\footnote{Note that their form of $\lambda_{16}$ has a typo, the $+i$ should be in the first column fifth row.} and defining the vector $\pi^a=(\pi_1,...,\pi_{35}),$ we can write out the contraction $\pi^a\lambda_a$ as
\begin{widetext}

\begin{equation}\label{matrix} \pi^a\lambda_a=
\left(
\begin{array}{cccccc}
 \pi _3+\frac{\pi _8}{\sqrt{3}}-\frac{\pi _{35}}{\sqrt{3}} & \pi _1-i \pi _2 & \pi _4-i \pi _5 & \pi _9-i \pi _{10} & \pi _{15}-i \pi _{16} & \pi _{21}-i \pi _{22} \\
 \pi _1+i \pi _2 & -\pi _3+\frac{\pi _8}{\sqrt{3}}-\frac{\pi _{35}}{\sqrt{3}} & \pi _6-i \pi _7 & \pi _{11}-i \pi _{12} & \pi _{17}-i \pi _{18} & \pi _{23}-i \pi _{24} \\
 \pi _4+i \pi _5 & \pi _6+i \pi _7 & -\frac{2 \pi _8}{\sqrt{3}}-\frac{\pi _{35}}{\sqrt{3}} & \pi _{13}-i \pi _{14} & \pi _{19}-i \pi _{20} & \pi _{25}-i \pi _{26} \\
 \pi _9+i \pi _{10} & \pi _{11}+i \pi _{12} & \pi _{13}+i \pi _{14} & \pi _{29}+\frac{\pi _{34}}{\sqrt{3}}+\frac{\pi _{35}}{\sqrt{3}} & \pi _{27}-i \pi _{28} & \pi _{30}-i \pi _{31} \\
 \pi _{15}+i \pi _{16} & \pi _{17}+i \pi _{18} & \pi _{19}+i \pi _{20} & \pi _{27}+i \pi _{28} & -\pi _{29}+\frac{\pi _{34}}{\sqrt{3}}+\frac{\pi _{35}}{\sqrt{3}} & \pi _{32}-i \pi _{33} \\
 \pi _{21}+i \pi _{22} & \pi _{23}+i \pi _{24} & \pi _{25}+i \pi _{26} & \pi _{30}+i \pi _{31} & \pi _{32}+i \pi _{33} & \frac{\pi _{35}}{\sqrt{3}}-\frac{2 \pi _{34}}{\sqrt{3}} \\
\end{array}
\right),
\end{equation}

\end{widetext}
where $\lambda_a=2T_a$.

In order to understand the quark composition of each $\pi$-axion, we imagine labeling the successive rows of Eq.~(\ref{matrix}) with the respective quark flavors $u,d,s,c,b,t$, while labeling the columns by the anti-quarks with the same ordering. When a $\pi$-axion appears in the $i^{\text{th}}$ column and $j^{\text{th}}$ row, its composition is identified to be that of quark-antiquark labeling of that entry. For example, note that $\pi_3$ appears in the first row/column with a positive sign, and the second row/column with a negative sign, so we deduce $\pi_3\sim u\bar{u}-d\bar{d}$. $\pi_3$ is therefore a real scalar and its own antiparticle. On the other hand, consider the $\pi$-axion $\pi_6+i\pi_7,$ which appears in the second column, third row of Eq.~(\ref{matrix}). Its composition is thus $d\bar{s}$ and since it is complex, its antiparticle is $\pi_6-i\pi_7$ of composition $\bar{d}s$. 

The $\pi$-axion field content is given in Table \ref{neutralmasstable}. Dark QCD where all six flavors of quark are light leads to 35 $\pi$-axions. Of these 17 are neutral and 18 are charged. Of the neutral $\pi$-axions, five are real pseudoscalars while the remaining 12 are grouped into six complex scalars. The charged $\pi$-axions are grouped into nine complex scalars. In summary, we have: five real, neutral states, six complex, neutral states, and nine complex charged states.

\begin{table*}
Spectrum of $\pi$-axions in Dark QCD\\
\begin{tabular}{c|c|c|c}

\hline
$\pi$-axion    & quark content       & mass squared ($m_{\pi^i} ^2$)     & charge [$\varepsilon$]    \\ \hline 
Real Neutral: & &\\ \hline
$\pi_3$ & $u\bar{u}-d\bar{d}$ & $(c_1+c_2)m_{\mathrm{I}} F_{\pi}$ & 0 \\
  $\pi_8$      &  $u\bar{u}+d\bar{d}-2s\bar{s}$    & $\left((c_1+c_2)m_{\mathrm{I}}+c_3m_{\mathrm{II}}\right ) F_{\pi}$  & 0\\
  $\pi_{29}$ & $c\bar{c}-b\bar{b}$ & $\left(c_4m_{\mathrm{II}}+c_5m_{\mathrm{III}}\right)F_{\pi}$ & 0\\
  $\pi_{34}$ & $c\bar{c}+b\bar{b}-2t\bar{t}$ & $\left( c_4m_{\mathrm{II}}+(c_5+c_6)m_{\mathrm{III}}\right) F_{\pi}$ & 0\\
$\pi_{35}$  & $-u\bar{u}-d\bar{d}-s\bar{s}+c\bar{c}+b\bar{b}+t\bar{t}$  & $\left((c_1+c_2)m_{\mathrm{I}}+(c_3+c_4)m_{\mathrm{II}}+(c_5+c_6)m_{\mathrm{III}}\right) F_{\pi}$   &0       \\ \hline Complex Neutral: & & \\ \hline
$\pi_6\pm i\pi_{7}$ &$d\bar{s}/\bar{d}s$ & $\left(c_2m_{\mathrm{I}}+c_3m_{\mathrm{II}}\right)F_{\pi}$ & 0\\
$\pi_9\pm i\pi_{10}$ &$u\bar{c}/\bar{u}c$ & $\left(c_1m_{\mathrm{I}}+c_4m_{\mathrm{II}}
\right)F_{\pi}$& 0 \\
$\pi_{17}\pm i\pi_{18}$ & $d\bar{b}/\bar{d}b$ & $\left(c_2m_{\mathrm{I}}+c_5m_{\mathrm{III}}\right)F_{\pi}$ & 0\\
$\pi_{19}\pm i\pi_{20}$ & $s\bar{b}/\bar{s}b$ & $\left(c_3m_{\mathrm{II}}+c_5m_{\mathrm{III}}\right)F_{\pi}$& 0 \\
$\pi_{21}\pm i\pi_{22}$ & $u\bar{t}/\bar{u}t$ & $\left(c_1m_{\mathrm{I}}+c_6m_{\mathrm{III}}\right)F_{\pi}$& 0 \\
$\pi_{30}\pm i\pi_{31}$ & $c\bar{t}/\bar{c}t$ & $\left(c_4m_{\mathrm{II}}+c_6m_{\mathrm{III}}\right)F_{\pi}$ & 0 \\ \hline Charged: & & \\ \hline 
$\pi_1\pm i\pi_2$ & $u\bar{d}/\bar{u}d$ & $(c_1+c_2)m_{\mathrm{I}}F_{\pi} + 2 \xi_1  (\te\varepsilon F_{\pi})^2$ & $\pm1$ \\
  $\pi_4\pm i\pi_5$ & $u\bar{s}/\bar{u}s$  & $ (c_1m_{\mathrm{I}}+c_3m_{\mathrm{II}})F_{\pi}+ 2 \xi_2 (\te\varepsilon F_{\pi})^2$ & $\pm1$   \\
  $\pi_{15}\pm i\pi_{16}$ & $u\bar{b}/\bar{u}b$  & $ ( c_1m_{\mathrm{I}}+c_5m_{\mathrm{III}})F_{\pi}+ 2 \xi_3 (\te\varepsilon F_{\pi})^2$ & $\pm1$ \\
   $\pi_{11}\pm i\pi_{12}$ & $d\bar{c}/\bar{d}c$  & $(c_2m_{\mathrm{I}}+c_4m_{\mathrm{III}})F_{\pi}+ 2 \xi_4 (\te\varepsilon F_{\pi})^2$ & $\mp1$ \\
 $\pi_{23}\pm i\pi_{24}$ & $d\bar{t}/\bar{d}t$  & $(c_2m_{\mathrm{I}}+c_6m_{\mathrm{III}})F_{\pi}+ 2 \xi_5 (\te\varepsilon F_{\pi})^2$ & $\mp1$ \\
 $\pi_{13}\pm i\pi_{14}$ & $s\bar{c}/\bar{s}c$  & $(c_3+c_4)m_{\mathrm{II}}F_{\pi}+ 2 \xi_6 (\te\varepsilon F_{\pi})^2 $ & $\mp1$ \\
 $\pi_{25}\pm i\pi_{26}$ & $s\bar{t}/\bar{s}t$  & $(c_3m_{\mathrm{II}}+c_6m_{\mathrm{III}})F_{\pi}+ 2 \xi_7 (\te\varepsilon F_{\pi})$ & $\mp1$ \\
  $\pi_{27}\pm i\pi_{28}$ & $c\bar{b}/\bar{c}b$  & $(c_4m_{\mathrm{II}}+c_5m_{\mathrm{III}})F_{\pi}+ 2 \xi_8 (\te\varepsilon F_{\pi})^2$ & $\pm1$ \\
 $\pi_{32}\pm i\pi_{33}$ & $b\bar{t}/\bar{b}t$  & $(c_5+c_6)m_{\mathrm{III}}F_{\pi}+ 2 \xi_9 (\te\varepsilon F_{\pi})^2 $ & $\mp1$\\
 \hline
\end{tabular}
\caption{{\bf Spectrum of $\pi$-axions}: Neutral real $\pi$-axions (top), complex neutral $\pi$-axions (middle), and charged $\pi$-axions (bottom). The first five $\pi$-axions are found along the diagonal of Eq.~(\ref{matrix}).
The other 12 $\pi$-axions are organized into 6 complex neutral fields.
\label{neutralmasstable} Finally, 18 of the $\pi$-axions are organized into 9 charged $\pi$-axions. Here we have approximated $F_{\pi}\sim \Lambda_{\rm dQCD}$ in expressing the mass formulae, and fixed $m_{\gamma'}>F_\pi$ in the charged $\pi$-axion mass relation Eq.~\eqref{eq:mpm}. }
\end{table*}

\section{\texorpdfstring{$\pi$}--axion-photon interactions}
\label{sec:interactions}

Experimental searches for axions are predicated on the characteristic interactions with the Standard Model photon and fermions. The $\pi$-axiverse is characterized by a mixture of parity-odd and parity-even couplings, as one might conventionally associate with an axion field and a dilaton field, respectively. 

We focus on interactions with Standard Model photons. We consider four different vertices such that $\mathcal{L}_{\rm int}=\sum_{i=1}^4\mathcal{L}_{\text{int}}^{(i)},$ where, momentarily omitting the sum over species,
\begin{eqnarray}\label{interaction1}
\mathcal{L}_{\text{int}}^{(1)}&&=\frac{\lambda_1}{2 F_{\pi}}\varepsilon^2(\pi^0)F_{\mu\nu}\tilde{F}^{\mu\nu},\\
\label{interaction2}
\mathcal{L}_{\text{int}}^{(2)}&&=\frac{\lambda_2}{2} \varepsilon^2 (\pi^{+})(\pi^{-}) A_\mu A^\mu, \\
\label{interaction3}
\mathcal{L}_{\text{int}}^{(3)}&&=\frac{\lambda_3}{2 \Lambda_3 ^2} \varepsilon^2 (\pi^{+})(\pi^{-}) F_{\mu\nu}F^{\mu\nu}, \\
\label{interaction4}
\mathcal{L}_{\text{int}}^{(4)}&&=\frac{\lambda_4}{2 \Lambda_4 ^2} \varepsilon^2(\pi_i)(\pi_j) F_{\mu\nu}F^{\mu\nu}. 
\end{eqnarray}
Here, ${\cal L}^{(1)}$ is the standard axion-photon coupling resulting from a triangle diagram, and ${\cal L}^{(2)}$ is the gauge covariant derivative of scalar QED, corresponding to the millicharges of the charged $\pi$-axions. 

\begin{figure}[h!]
\centering
    \includegraphics[scale=0.7]{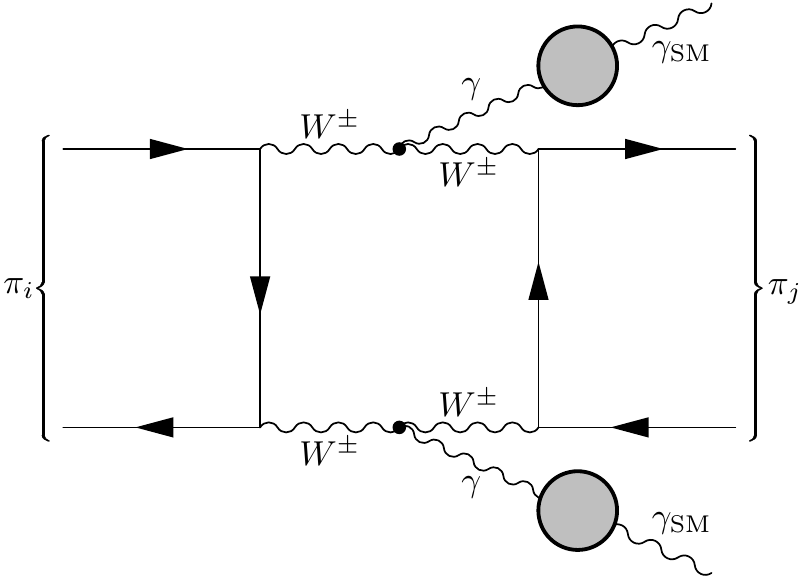}
    \caption{\label{fig:mixing_box}  A diagram illustrating a possible UV process contributing to the $\pi\pi\gamma\gamma$ effective vertices. All particles are presumed to reside in the dark SM unless explicitly noted. The grey blobs refer to kinetic mixing between the dark and visible photons.}
\end{figure}

Meanwhile ${\cal L}^{(3)}$ and ${\cal L}^{(4)}$ arise  as effective field operators arising from integrating out the heavy degrees of freedom in the dark Standard Model, the former interaction describing all charged $\pi$-axions and the latter describing all neutral (real and complex) $\pi$-axions. As in the visible SM, the weak bosons are electrically charged and can couple to the photon as a result. Therefore at high energies, two neutral $\pi$-axions can interact with photons through a loop process involving weak bosons as in $K-\bar{K}$ mixing in the visible SM. The photons could be emitted by the dark $W^{\pm}$ or possibly by the off-shell quarks.  A diagram illustrating this process is shown in figure Fig.~(\ref{fig:mixing_box}). From this we can estimate the energy scales $\Lambda_{3,4}$ as simply ${\Lambda_{3,4} } \sim m_{W} = g v$.

The interactions (\ref{interaction2})-(\ref{interaction4}) can describe pair creation, annihilation, or decay processes if the $\pi$-axions are two different species of the same charge. The heaviest states can proceed down a decay chain, however, the lightest complex $\pi$-axion (charged or neutral) cannot decay further and are completely stable. $\mathcal{L}^{(1)}_{\text{int}}$ is analogous to the Standard Model neutral pion interaction, which corresponds to a decay width given by 
\begin{equation}
\Gamma_{(1)}=\varepsilon^4\frac{\lambda_1^2\alpha_e^2}{64\pi^3}\frac{m_i^3}{F_\pi^2}.
\end{equation}
Demanding the real, neutral $\pi$-axions be cosmologically stable can place approximate bounds on the parameters in our model, which we will discuss further in the following section.

\section{ Cosmology of the \texorpdfstring{$\pi$}--axiverse}
\label{sec:cosmo}

There are many possibilities for the cosmological evolution of the $\pi$-axiverse \cite{Preskill:1982cy,Turner:1990uz,Hertzberg:2008wr}. The $\pi$-axions can range in mass across the whole spectrum conventionally associated with axion-like particles, including the benchmark mass for experiments such as ADMX, $m\sim 10^{-5}$ eV, `fuzzy dark matter' \cite{Hu:2000ke} with $m\sim 10^{-21}$ eV (see \cite{Maleknejad:2022gyf} for an analysis of this case), and much lighter particles which could be a sub-component of dark matter (see \cite{Hlozek:2014lca,Lague:2021frh}) or else contribute to dark energy.

We focus on the scenario in which the dark QCD phase transition occurs before cosmic inflation, i.e., we assume that
\begin{equation}
    F_{\pi} > H_{\text{inf}} .
\end{equation}

The relic density in $\pi$-axions can then be generated by the misalignment mechanism. Similar to the conventional cosmic history of axions \cite{Marsh:2015xka}, the $\pi$-axions are initialized after the symmetry breaking with a random value (`misalignment'), corresponding to the angular excitations of the system in the directions of the broken generators of the $SU(N_f)_{\rm L}\times SU(N_f)_{\rm R}$ flavor symmetry, which is spontaneously broken by the dQCD condensate. The subsequent energy density is determined by the ($\pi$-)axion mass and initial misalignment. In the present context, the $\pi$-axions acquire masses due to the dark electroweak phase transition, at which point the fields acquire a periodic potential. Since the fields have no method of selecting their position prior to the emergence of the potential, they may be misaligned with its minima and thus have differing energies with respect to each other depending on the degree of misalignment. The relative misalignment angles are a free parameter of this class of models.

Similar to a conventional axion, this scenario is constrained by cosmic microwave background isocurvature, which imposes an upper bound on the energy scale of inflation  $H_{\rm inf} \lesssim 10^{10} (F_{\pi}/M_{\text{pl}}) {\rm GeV}$ \cite{Marsh:2015xka}. To simplify the analysis we further assume that the temperature of the Standard Model plasma is below the dark QCD energy scale at all times in the radiation dominated epoch, i.e., 
\begin{equation}
\label{eq:Tre}
    \Lambda_{\rm dQCD}\sim F_{\pi} > T_{\rm max},
\end{equation}
where $T_{\rm max}$ is the maximal temperature of the Standard Model plasma, which is at most $\sim \sqrt{H_{\rm infl}M_{\rm pl}}$. This  prevents any freeze-in or freeze-out production of heavy states in the dark QCD sector, allowing us to focus solely on the light degrees of freedom.

We further demand that the $\pi$-axions are cosmologically stable and will not decay, e.g., to dark or visible sector photons. The neutral complex and charged $\pi$-axions are stable in our model, whereas the neutral  can decay, into either dark photons or Standard Model photons. Decay into Standard Model photons gives the neutral scalar $\pi$-axions a lifetime,
 \begin{equation}
 \label{eq:taupi}
      \tau_\pi\approx( 4\times 10^7)\frac{F_\pi^2}{\varepsilon^4m_\pi^3},
 \end{equation}
 while decay into dark photons is can be divided into two cases: (1) the massless dark photon, the same as above but $\varepsilon=1$, and (2) the massive case, where decay is kinematically forbidden.

Constraints on decaying dark matter impose that the lifetime be at least $\gtrsim {\cal O}(10)$ times longer than the age of the universe $\tau_U = H_0 ^{-1}$ (see e.g. \cite{Simon:2022ftd} for a detailed analysis). This is easily satisfied in our model, wherein the large hierarchy $F_{\pi}/m_{\pi} \gg 1$ is a feature of the model \textit{a priori}. For example, a decay constant $F_{\pi}=10^{11}$ GeV and mass $m_{\pi}=10^{-5}$ eV, and unit $\varepsilon$, yields $\tau_\pi\sim 10^{29} H_0 ^{-1}$, well in agreement with observational constraints. Larger decay constants and smaller masses, or smaller millicharge, will only increase the neutral $\pi$-axion lifetime.  It additionally follows that neutral $\pi$-axions can be a stable axion-like dark matter candidate even in the case that the dark photon is massless, since in this case the lifetime is simply Eq.~\eqref{eq:taupi} with $\varepsilon=1$.

A late-time relic density of $\pi$-axions can be produced by the conventional misalignment mechanism (see \cite{Marsh:2015xka} for a review). The total relic density in $\pi$-axions produced via misalignment is given by
\begin{equation}\label{misa}
    \Omega_{\pi} = \frac{1}{6} (9 \Omega_r)^{3/4} \frac{F_\pi^2}{M_{\text{Pl}}^2} \displaystyle \sum _{i} \left(\frac{m_{\pi_i}}{H_0} \right)^{1/2}  \theta_{\pi_i} ^2,
\end{equation}
where the sum is over all light stable $\pi$-axion fields, and we assume $m_{\pi^i}>10^{-28}$ eV for simplicity (see \cite{Marsh:2015xka} for the relic density when $m_{\pi^i}<10^{-28}$ eV). The sum is fixed by observation, $\Omega_{\rm DM} h^2 = 0.12$ \cite{Planck:2018vyg}. The relative contribution of each $\pi$-axion to the relic density depends on the mass spectrum of $\pi$-axions, which in turn encodes the dark quark masses, dark QCD scale ($\Lambda_{\rm dQCD}\sim F_{\pi}$), and the millicharge $\varepsilon$.

Analogous to conventional axion models, enforcing that the relic density match observation leads to a relation between the dark matter mass and decay constant, however, in this case there is a {\it spectrum
} of masses and a set of random initial misalignments. We detail the features of this spectrum in the subsections to come. 

 We are interested in sub-eV $\pi$-axion masses, so that at least one of the fields can evolve as a coherent scalar that survives until the present day. Demanding that the $\pi$-axions constitute an ${\cal O}(1)$ fraction of the observed dark matter density, then enforces a lower bound on the decay constant,
\begin{equation}\label{FpiLowerBound}
    F_\pi > 8.8 \times 10^{10} {\rm GeV} .
\end{equation}
From this we deduce an upper bound mass range for the mass of the dark quarks for the lightest  to be sub-eV, 
\begin{equation}\label{mass-ULP}
    m_{q} \lesssim 5 \times 10^{-20} {\rm eV}.
\end{equation}
The details of the spectrum of $\pi$-axions is dictated by a combination of the quark masses and the millicharge $\varepsilon$. We investigate each of these in turn. 

\subsection{ Parameter Space for Dark Matter}

 \subsubsection{  Quark Masses.}\label{subsec:masshierarchies}

We first consider the case wherein the millicharge makes a subdominant contribution to the $\pi$-axion masses, and we consider $\varepsilon=0$ for simplicity.

In the case that the three generations are degenerate, $m_{\text{I}}=m_{\text{II}}=m_{\text{III}}=m_q$ , all 35 $\pi$-axions have masses around $m_\pi \sim \sqrt{m_q F_{\pi}}$, with a distribution set by the $c_i$. In the completely degenerate case where all $c_i$ are equal to one, then there are solely three unique masses:
\begin{equation}\label{threedegenmasses}
    \begin{split}
        m_2&\sim\sqrt{2m_qF_\pi}, \\
        m_3&\sim\sqrt{3m_qF_\pi}, \\
        m_6&\sim \sqrt{6m_qF_\pi}.
    \end{split}
\end{equation}
All of the complex $\pi$-axions as well as, for example $\pi_3$ and $\pi_{29}$, are composed of two unique quarks with mass $m_2$. The two real $\pi$-axions $\pi_8$ and $\pi_{29}$ are composed of three quarks and have mass $m_3$, while $\pi_{35}$ is composed of six quark flavors and is the most massive, with its mass approximately near $m_6$.

If we relax the condition  that $m_{\rm I,II,III}$ are equal, then the spectrum splits into three distinct regions, with $\pi$-axion masses clustered around each of $\sqrt{m_{\rm I} F_\pi}$,  $\sqrt{m_{\rm II}F_\pi}$, and $\sqrt{m_{\rm III}F_\pi}.$ We note that, a priori, the mass hierarchies are inputs of the model and thus may be set arbitrarily, but not so far apart as to spoil the flavor symmetry of the quarks. 

\begin{figure*} 
\includegraphics[width = 0.49\textwidth]{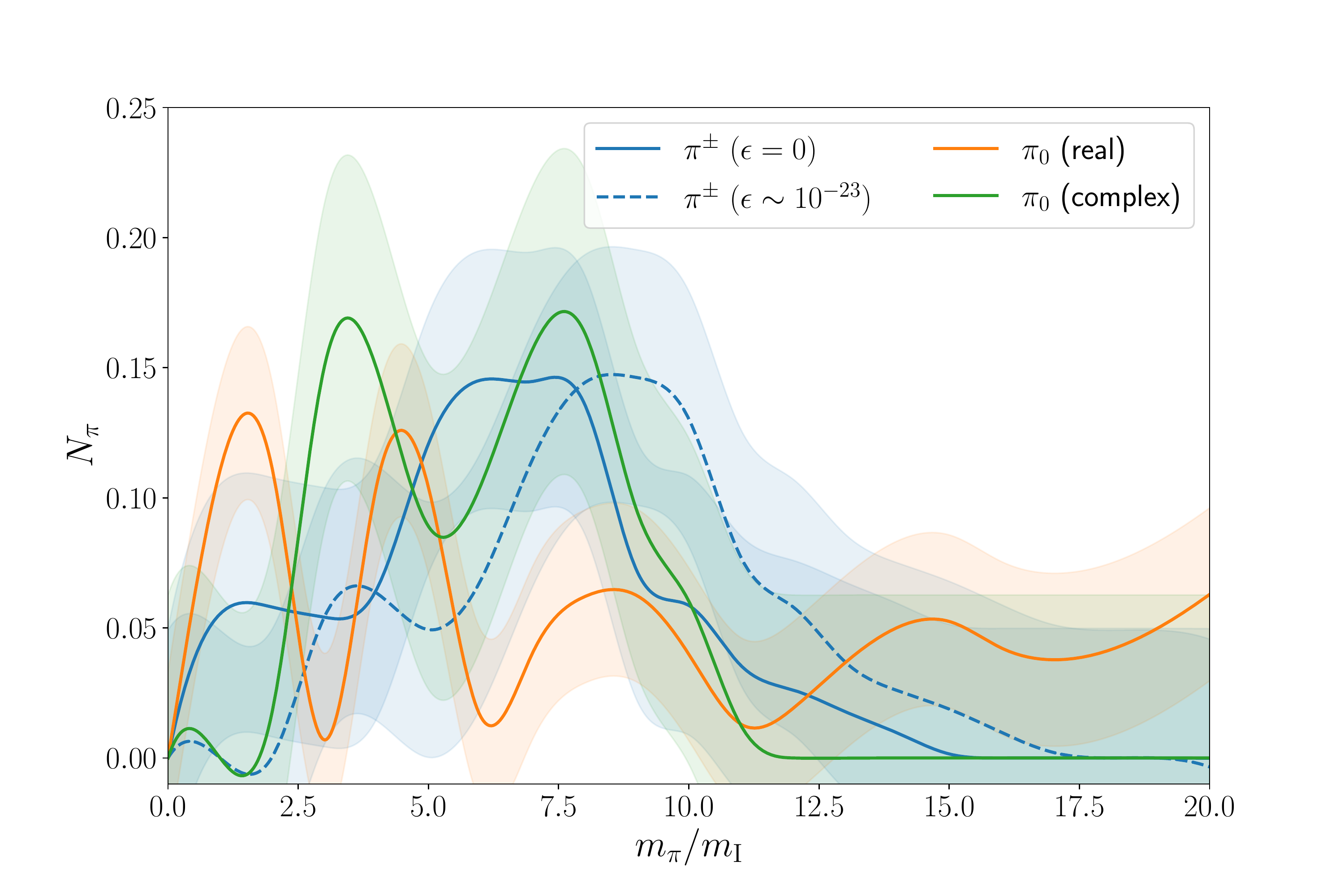} 
\includegraphics[width =0.49\textwidth]{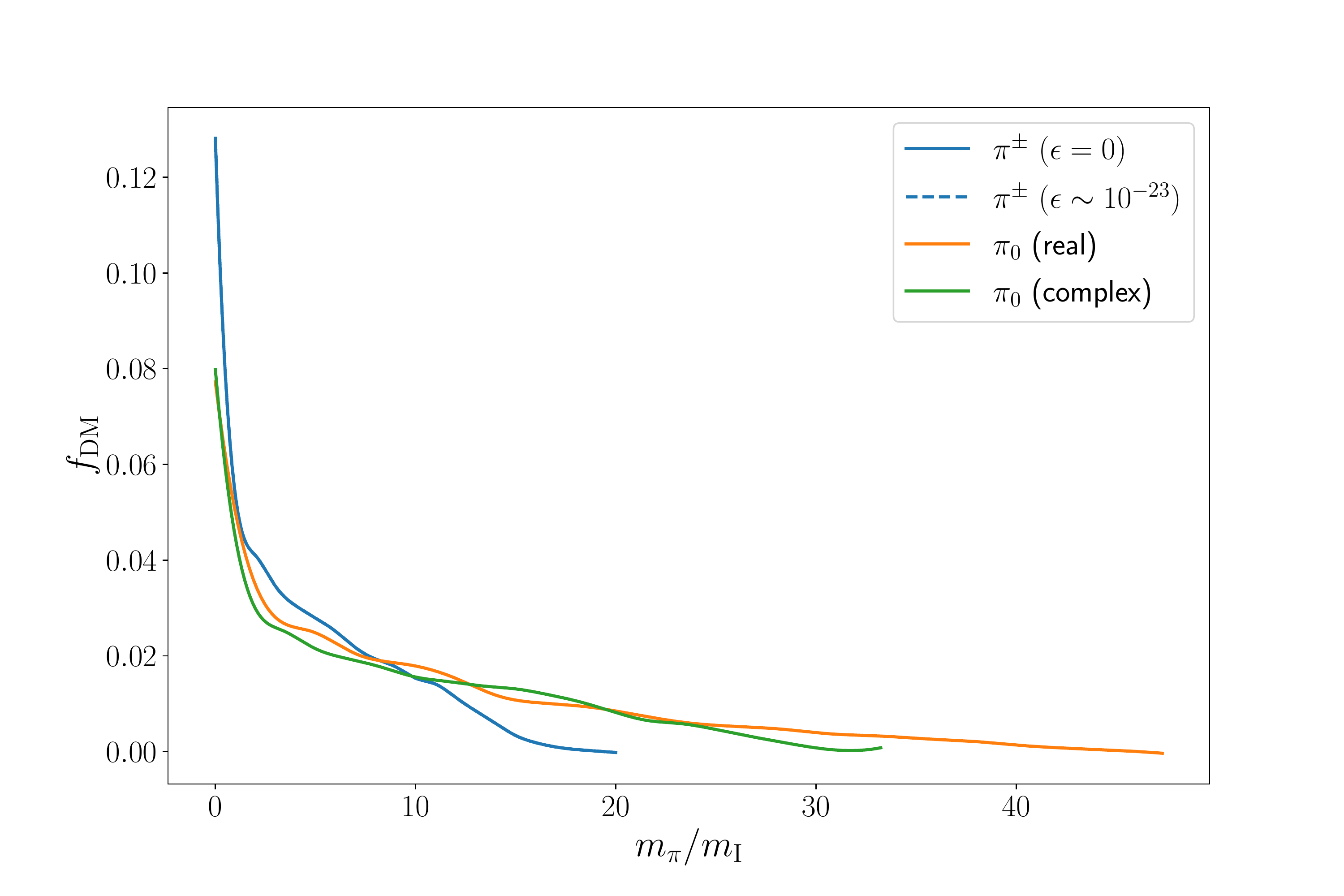} 
\caption{\label{fig:spectra}
Statistical mean (solid) and $\pm1\sigma$ (shaded band) $\pi$-axion mass spectra (left) and mass distribution of fractional contribution to DM relic density (right) for $m_{\rm I}$, $m_{\rm II}$, $m_{\rm III}$ given by  $1$, $3$, and $6 \times10^{-5}\text{ eV}$ respectively.  Orange is the real pseudoscalar neutral $\pi$-axions, green is complex neutral $\pi$-axions, and blue is charged $\pi$-axions. The charged pions are shown with vanishing millicharge ($\varepsilon = 0$) and millicharge $\varepsilon\sim10^{-23}$ with the solid and dashed blue lines, respectively. The dark matter relic density is nearly evenly split between the three classes of $\pi$-axion, and the distribution features a heavy tail towards larger $\pi$-axion masses. Here we assume a heavy dark photon, $m_{\gamma'}>F_{\pi}$.
}
\end{figure*}

To understand the mass spectrum, we compute the statistical average spectrum, shown in Fig. \ref{fig:spectra}. We consider a scenario where $m_{\mathrm{I}}<m_{\mathrm{II}}<m_{\mathrm{III}}$, with $m_{\rm I}$, $m_{\rm II}$, $m_{\rm III}$ given by  $1$, $3$, and $6 \times10^{-5}\text{ eV}$ respectively, and with the splitting coefficients $c_i\sim1$ drawn from uniform distributions, $c_i = [0.7,1.3]$. In this case, the $\pi$-axion masses are spread out across the spectrum, with larger masses possible; note that the charged and neutral complex distributions mostly mirror each other. Using Eq.~(\ref{misa}), it is possible to compute a breakdown of the dark matter densities in terms of each species of $\pi$-axion; as expected, since the mis-alignment angles are a priori completely uncorrelated, then no one particular type dominates, as seen in Fig. \ref{fig:spectra}.

\subsubsection{Dark Photon Mass}

If $m_{\gamma'}< F_{\pi}$, then the electromagnetic contribution to the charged pion mass is dominated by the dark photon, generating a large ($\approx F_{\pi}$) mass for the charged pions, see Eq.~\eqref{eq:mpm}. On the other hand, for $m_{\gamma'} \gg F_{\pi}$, the charged pion mass is dominated by the visible SM photon and the charged $\pi$-axion mass is $\varepsilon$-suppressed as $m_{\pm}\sim\varepsilon F_{\pi}$.  In the former case, the charged $\pi$-axions will decay and annihilate to dark photons before the onset of inflation, leaving no trace in the late universe. In the  latter case, the decay of the charged $\pi$-axions can be delayed, but since the decay channel is to SM photons, there is again no dark radiation constraint. As stated previously, we will proceed focusing on the case $m_{\gamma'}>F_\pi$.

\subsubsection{Millicharge $\varepsilon$}

When $\varepsilon=0$, the charged $\pi$-axion  masses are comparable in size with the neutral $\pi$-axions. In this case, the relic density is distributed evenly across all  species, according to the distribution of masses and initial misalignments. 

When $\varepsilon$ is non-zero, the electromagnetic contribution to the charged  masses, $\sim e\varepsilon F_{\pi}$, can become relevant.  The charged $\pi$-axions will be coherent scalars in the late universe, i.e. have $m<$ eV, only if the millicharge is below an upper bound given by $\varepsilon  < \frac{1~ {\rm eV}}{F_{\pi}}. $
Given that $F_{\pi}\gtrsim 10^{11}$ GeV in order for sub-eV particles to have any significant relic density, this translates to a bound $\varepsilon  \lesssim 10^{-20} 
$ in order for the charged $\pi$-axions to enjoy an axion-like cosmological history and contribute to the DM relic density via their misalignment production. This bound scales with $F_{\pi}$ as $10^{11} {\rm GeV}/{F_{\pi}}$. 

Non-zero $\varepsilon$ also allows the neutral, real $\pi$-axions to decay to two photons, via the conventional axion channel. Nonetheless -- the neutral $\pi$-axions remains cosmologically stable even for ${\cal O}(1)$ millicharge:  they have lifetime longer than the age of the universe  if, 
\begin{equation}
   10^{-34} \varepsilon^4   \left( \frac{m_\pi}{ 10^{-5} \, {\rm eV}}\right)^3 \left(\frac{10^{13} {\, \rm  GeV}}{F_{\pi}} \right)^2\lesssim 1.
\end{equation}
From this we conclude that sub-eV neutral $\pi$-axions are a stable dark matter candidate for the full range of $\varepsilon$ and $F_{\pi}$ we consider.

For $\varepsilon \gg 10^{-20}$, the charged $\pi$-axions will be heavy ($m_\pm \sim \epsilon F_{\pi}$, see Eq.~\eqref{eq:mpm}), and cease to evolve as a coherent scalar field already in the very early universe. In this case one expects a negligible relic density in charged $\pi$-axions via misalignment.

 Assuming the range of $F_{\pi}$ given by Eq.~\eqref{eq:Tre}, this regime of $\varepsilon$ can be further subdivided into two regimes: (1) for $\varepsilon > \sqrt{H_{\rm inf}/M_{pl}}$, then $m_{\pm} > H_{\rm inf}$ and the initial misalignment of $\pi$-axions  will decay before the onset of inflation, leaving no discernible trace in the post-inflationary universe; (2) for $\varepsilon < \sqrt{H_{\rm inf}/M_{pl}}$, the decay of the coherent charged $\pi$-axion fields will occur after inflation, and the resulting non-relativistic charged $\pi$-axions will scatter with SM particles and annihilate to SM photons with a rate $\propto \varepsilon^2$. For a detailed discussion of constraints on $\varepsilon$ from charged $\pi$-axions we refer the reader to \cite{Maleknejad:2022gyf}, and note here that the values of $\varepsilon$ needed to engineer charged $\pi$-axions that are low mass enough be cosmologically relevant are so small as  to easily satisfy constraints.

Finally, we can consider $\varepsilon \lesssim {\cal O}(1)$. In this case, all of the charged $\pi$-axions will be extremely massive $m\sim F_{\pi}$. While there are stringent bounds on electrically charged dark matter \cite{Davidson:2000hf,Bogorad:2021uew,Alexander:2020wpm, Plestid:2020kdm}, these bounds are easily satisfied for heavy masses (e.g. $m\sim F_{\pi}$) and particles with zero relic density. It is therefore important to check the relic density produced by means other than misalignment, such as by freeze-in or freeze-out processes \cite{Hall:2009bx,Elahi:2014fsa,Chu:2011be,Du:2021jcj}. We will consider both of these scenarios in turn in the subsequent section.

In summary, the  $\varepsilon$-dependence of the model is split into two regimes: $\varepsilon \ll 10^{-20}$ and $\varepsilon \gg 10^{-20}$, corresponding to the presence or absence of a relic density of charged $\pi$-axions produced via misalignment. For $\varepsilon \gg 10^{-20}$, the dark matter relic density is comprised of real and complex neutral $\pi$-axions. 

\subsection{Freeze-Out and Freeze-In  of Charged \texorpdfstring{$\pi$}--axions}

As one considers light charged $\pi$-axions, $m_{\pm} \ll F_{\pi}$, there is also the possibility of their freeze-in or freeze-out production. A sufficient condition to satisfy  bounds on millicharged matter \cite{Plestid:2020kdm}, is that the relic density generated through freeze-in or freeze-out is vanishingly small. We consider these in turn.

\textit{Freeze-in}: A detailed derivation of the relic density of heavy millicharged dark matter through freeze-in QED processes was given by one of the present authors in \cite{Maleknejad:2022gyf}. The main result is expressed in terms of the equation of state during reheating  $w_{\rm re}=p/\rho$  as
\begin{equation}\label{Odm1}
    \Omega_{\pi^\pm} h^2\simeq (5\times 10^{-3})\mathcal{A}(w_{\rm re})\varepsilon^4\frac{\text{exp}[10(3-\alpha\sqrt{\beta})]}{\beta^{4/(1+w)-1/2}},
\end{equation}
where $\mathcal{A}(w)=\frac{(10\pi^2)^{-w/(1+w)}}{2(1+w)},$ $\alpha=\frac{m_{\pi^\pm}}{\sqrt{H_{\text{inf}}M_{\text{pl}}}}$, and $\beta=\frac{\sqrt{H_{\text{inf}}M_{\text{pl}}}}{T_{\text{reh}}}$. 
To guarantee a negligible relic density $\Omega_{\pi^\pm}h^2\ll 0.12$, one may focus on the limiting case that $\varepsilon\sim \mathcal{O}(1)$. As an example, consider $H_{\rm infl}=10^{10}$ GeV and reheating with $w=0$, with a maximum temperature $T_{\rm re} \sim 10^{9}$ GeV. The relic density in this case reads,
\begin{equation}
    \Omega_{\pi^\pm}h^2\simeq  4 \times 10^{-8}\varepsilon^4e^{ - {\varepsilon F_\pi}/( {10^{11}\text{ GeV}})},
\end{equation}
implying that the relic density, even for $\varepsilon=1$, is at most $\sim 10^{-7}$ for $F_{\pi}>10^{11}$ GeV, and thus charged $\pi$-axions produced via freeze-in constitute an extremely small fraction of the total dark matter. From this we infer that freeze-in production of charged $\pi$-axions, for the axion dark matter-like parameter regime $F_{\pi} > 10^{11}$ GeV, always yields a vastly subdominant contribution to the dark matter density.

\textit{Freeze-out}: 
Precluding freeze-out production may be enforced by demanding that the charged $\pi$-axions are never in thermal equilibrium. The following bound on the millicharge follows:
\begin{equation}\label{freezeoutbound1}
    \varepsilon\ll (4\times 10^2)\Big(\frac{m_{\pi^{\pm}}}{M_{\text{pl}}}\Big)^{1/2}\Big(\frac{m_{\pi^\pm}}{T_{\text{max}}}\Big)^{1/4}e^{\frac{m_{\pi^\pm}}{2T_{\text{max}}}}.
\end{equation}
In Eq.~(\ref{freezeoutbound1}), $T_{\text{max}}$ is the maximum temperature achieved during reheating \cite{Chung:1998rq}. We can equivalently write Eq.~(\ref{freezeoutbound1}) as the transcendental relationship
\begin{eqnarray}\label{freezeoutbound2}
    \varepsilon\ll (6.5\times 10^8)\Big(\frac{F_\pi}{M_{\text{pl}}}\Big)^2\Big(\frac{F_\pi}{T_{\text{max}}}\Big)e^{2e\varepsilon F_\pi/T_{\text{max}}},
\end{eqnarray}
which again takes the assumption that $m_{\pi^\pm}\sim e\varepsilon F_\pi$. Consider, for example, a GUT scale decay constant, $F_{\pi}=10^{15}$ GeV, and a universe that reheats to a temperature well below the GUT scale, e.g. $T_{\rm max}\sim 10^{12} $ GeV. In this case the bound on $\varepsilon$ reads $\varepsilon \ll 10^{4} e^{ 100\varepsilon}$. This is satisfied in the full range of $\varepsilon$ from 0 to 1. Thus we conclude that charged $\pi$-axions are too heavy to be in thermal equilibrium, even with unit electric charge.

\subsection{ Summary of Dark Matter Scenarios}

The possibilities for dark matter can be summarized as follows:
\begin{itemize}
    \item Vanishing millicharge ($\varepsilon =0$) and degenerate quark generation mass scale ($m_{\rm I}=m_{\rm II}=m_{\rm III}=m_q$): dark matter is a combination of all 35 $\pi$-axions, with masses distributed near the shared mass scale $m_\pi \sim \sqrt{m_q F_{\pi}}$ .
    \item $\varepsilon=0$ with splitting between quark generations: $m_{\rm I}<m_{\rm II}<m_{\rm III}$. Dark matter is a mixture of 35 $\pi$-axions, with masses clustered around the three mass scales.
    \item $0< \varepsilon < 10^{-20} $: the charged $\pi$-axions get a mass correction $m_{\pi^\pm}\sim \varepsilon F_{\pi}$ and can become much heavier than the neutral (real and complex) $\pi$-axions, adding additional structure to both the degenerate and non-degenerate generation cases.
    \item $10^{-20}\ll \varepsilon \lesssim {\cal O}(1)$: the charged $\pi$-axions are heavy and do not contribute to the late-time relic density. Dark matter is comprised of the real and complex neutral $\pi$-axions, with masses distributed according to $m_{\rm I}$, $m_{\rm II}$, $m_{\rm III}$. 
\end{itemize}

\section{\texorpdfstring{$\pi$}--axion Dark Matter Detection Prospects}
\label{sec:detection}

The search for ultralight dark matter candidates such as axions is a multifaceted and growing field, see e.g.~\cite{Antypas:2022asj} for a recent review. These approaches can be directly applied to $\pi$-axions, targeted at interactions of $\pi$-axions with the Standard Model that are generated both by the millicharge portal and by the heavy electroweak sector. For concreteness, here we focus on detection via photons, utilizing the interactions Eqs.~\eqref{interaction1}-\eqref{interaction4}.

\subsection{  Direct Detection via Parity-Odd Portal}

The long-standing experimental approach to axion direct detection has been to exploit the axion-photon coupling predicted by anomaly considerations, namely the coupling of the pseudoscalar axion to the parity-odd combination of electromagnetic field strengths $F\tilde{F}\equiv \varepsilon_{ \mu \nu \sigma \rho}F^{\mu \nu}F^{\sigma \rho}$.

The effective interaction following from the triangle anomaly diagram is given by,
\begin{equation}\label{L4}
    \mathcal{L}_{\text{int}}^{(1)}\sim\frac{\lambda_1}{2F_\pi}\alpha_e\varepsilon^2(\pi^0)F_{\mu\nu}\tilde{F}^{\mu\nu},
\end{equation}
which is of the same form as the standard axion-photon interaction $\mathcal{L}_{a \gamma \gamma} = g_{a \gamma \gamma} a F \tilde{F}/4$ \cite{Wilczek:1987mv, Zyla:2020zbs},
where in our case we have
\begin{equation}\label{gaxion}
    g_{\pi\gamma\gamma} \sim\frac{\alpha_e \varepsilon^2}{F_\pi}\lambda_1.
\end{equation}
The parameter $\lambda_1$ is analogous to the anomaly coefficient appearing in conventional axion models, and depending on the precise details of the dark Standard Model, can be ${\cal O}(1)-{\cal O}(10)$. The scaling with $\varepsilon^2$ reflects the millicharge portal of the dark $\pi$-axions to the Standard Model photon. 

\begin{figure*}
\centering
    \includegraphics[scale=0.45]{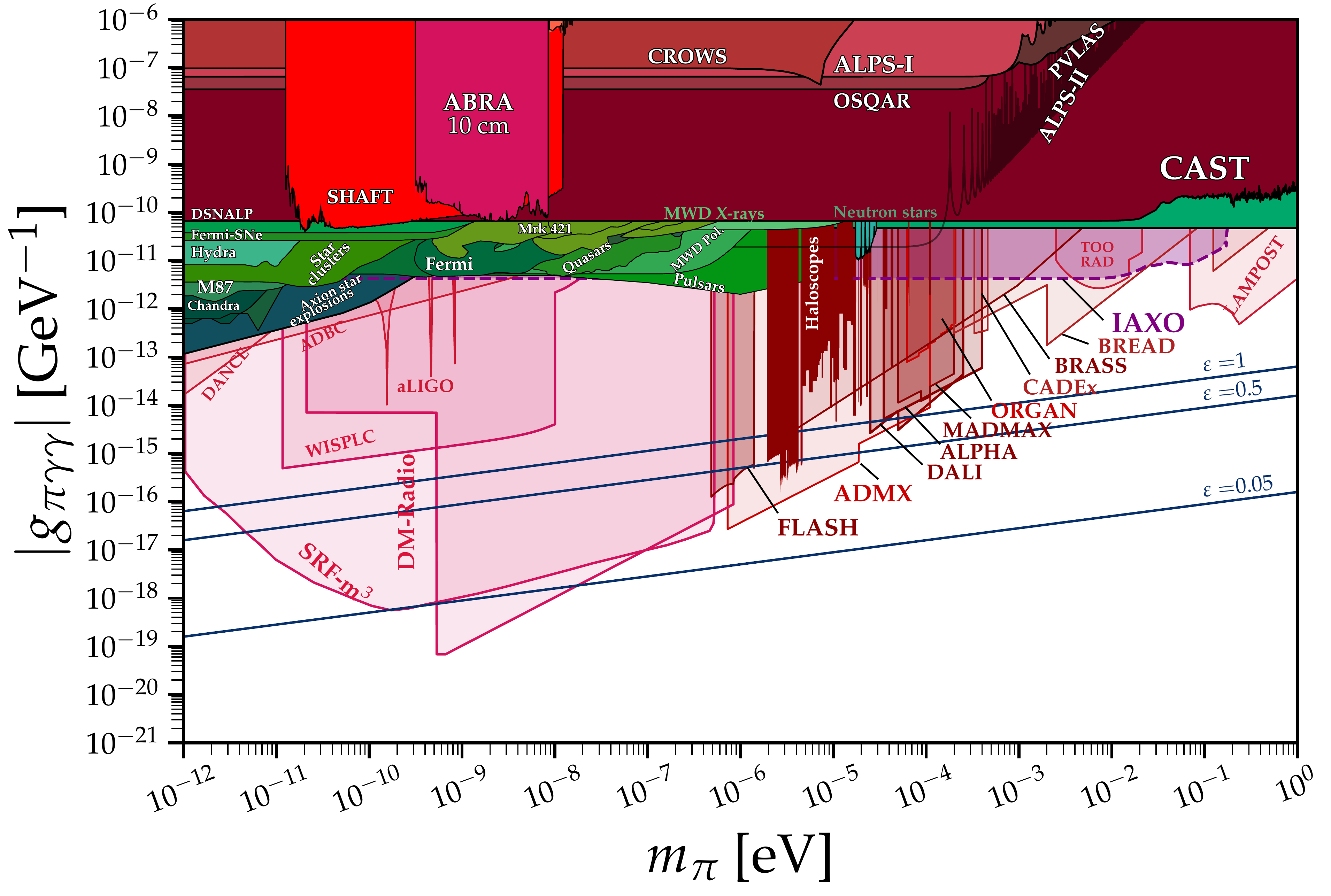}
    \caption{\label{fig:Axion_limits} 
    An exclusion figure illustrating the parameter space limits for a simplified limit of the $\pi$-axiverse model, described in the text. For this scenario, the five neutral $\pi$-axions are assumed to have a common mass scale and equal misalignment angles $\theta_i = 1$, and to constitute all of the dark matter density. The result is that $g_{\pi\gamma\gamma}\propto\lambda\epsilon^2\theta m_{\pi}^{\frac{1}{2}}$, a function of the common mass scale $m_{\pi}$ and depends quadratically on the millicharge $\varepsilon$ and linearly on the coupling $\lambda_1$, taken to be $\lambda_1 = 1$, and misalignment angle $\theta_i$, taken to be $\theta_i = 1$. The values of $\varepsilon = 1, 0.5$ and $0.05$ are shown as solid lines. Figure modified from \cite{AxionLimits}.
    }
\end{figure*}

This coupling is the focus of many axion dark matter haloscope experiments both in the $\mu$eV and above mass range and the sub-$\mu$eV range (see \cite{Adams:2022pbo} for a thorough overview of the ongoing efforts). Of note is the ADMX collaboration \cite{ADMX:2020ote}, whose attention is to the 1-20 $\mu$eV axion mass range (see also \cite{Lawson:2019brd,ALPHA:2022rxj,Shaposhnikov:2023pdj} for recent proposals of plasma haloscopes). Their most recent report bounds the axion decay constant at $g_{a\gamma\gamma}\lesssim 10^{-13}$ GeV$^{-1}$ for the narrow mass range $m_a\sim 19.83-19.845$ $\mu$eV \cite{ADMX:2021mio}, while \cite{ADMX:2021nhd} bounds $g_{a\gamma\gamma}\lesssim 10^{-15}$ GeV$^{-1}$ in the $m_a\sim 3-4$ $\mu$eV range. The HAYSTACK collaboration's phase 1  \cite{HAYSTAC:2018rwy} similarly reported the bound $g_{a\gamma\gamma}\lesssim10^{-15}$ GeV$^{-1}$ for $m_a\sim 23.15-24 \ \mu$eV. There are various other probes that place an upper bound on $g_{a\gamma\gamma}$ across the larger mass range of $m_a\sim 10^{-14}-10^{-5}$ eV, see e.g. the collection of constraint plots publicly available at \cite{AxionLimits}. Throughout that mass range, smaller values of $g_{a\gamma\gamma}$ are consistent with data. Due to the millicharge parameter $\varepsilon$ in Eq.~\eqref{gaxion}, our model can easily accommodate such small couplings.

In our model, there are five different $\pi$-axions that participate in the interaction, mainly $\pi_3, \ \pi_8, \ \pi_{29}, \ \pi_{34},$ and $\pi_{35}$ (see Tab. \ref{neutralmasstable}), each of which can have a size-able relic density produced via misalignment.  ADMX therefore has the potential to see five distinct resonances corresponding to the five different masses. In the completely degenerate case where $m_{\text{I}}=m_{\text{II}}=m_{\text{III}}\equiv m$, and all the $c_i$ are equal, then $m_{\pi_3}=m_{\pi_{29}}\sim m_2,$ $m_{\pi_8}=m_{\pi_{34}}\sim m_3$, and $m_{\pi_{35}}\sim m_6$ where $m_2,m_3$ and $m_6$ are defined in Eq.~\eqref{threedegenmasses}. This implies that exactly three resonances would be observed at approximately $1.4\sqrt{mF_\pi}, \ 1.7\sqrt{mF_\pi}$, and $2.45\sqrt{mF_\pi}$; these may be more or less degenerate depending on the resolution of the detector. In the more general case of fully non-degenerate quark masses, there will be exactly five peaks which potentially have greater than $\mathcal{O}(1)$ separation.

As an illustrative example, proposed limits arising from this scenario are shown in Fig.~\ref{fig:Axion_limits}, a modified version of the figure in \cite{AxionLimits}. Here we consider only the constraints on a single signal, without considering the integrated effect of 5 correlated signals of the neutral $\pi$-axions. The expected couplings as a function of $\varepsilon$ are shown as solid lines, with the shading indicating the range of possible observations due to the spread of values. The fully-degenerate scenario was implemented using Eq. (\ref{misa}), assuming the neutral $\pi$-axions constitute the entirety of the DM so that $\Omega_{\pi}h^2 = \Omega_{\rm DM}h^2 = 0.12$ and that all misalignment angles are degenerate and equal. This allows the extraction of $F_{\pi}$ as a function of $m_\pi$, the common $\pi$-axion mass scale in this scenario, and thus the computation of $g_{\pi\gamma\gamma}$ as a function of the mass and millicharge. In general, $g_{\pi\gamma\gamma}\propto\lambda\epsilon^2\theta_i m_{\pi}^{\frac{1}{2}}$. For concreteness in Fig.~\ref{fig:Axion_limits}, we assume $\lambda_1 = 1$, $\theta_i = 1$ and display various curves for $\varepsilon$.

Parts of the parameter space of this scenario are already excluded by  experiments such as ADMX, especially for greater values of the millicharge such as $\varepsilon \sim 1$; smaller values are mostly beyond experimental reach for many present and future experiments. DM-Radio can probe some regions of this parameter space, particularly for $m_{\pi}\sim10^{-10}-10^{-7}$. Reducing $\lambda_1$ or $\varepsilon$, or decreasing the misalignment angle $\theta_i$ moves this scenario out of almost all present experimental exclusions. It is worth emphasizing that this benchmark scenario is simplified, and that a more thorough treatment would require statistical averages over distributions of parameters.

\subsection{Atomic Oscillations via the Parity-Even Portal}

An alternative, more recent, approach to ultralight dark matter detection is to exploit the possibility of coupling to parity-even combinations of fields (see e.g. ~\cite{Sakurai:2021ipp}), such as couplings to $F_{\mu \nu}F^{\mu \nu}$. We consider the parity-even interactions of the $\pi$-axions
given by Eqs.~\eqref{interaction2}, \eqref{interaction3}, and \eqref{interaction4}.
All of the $\pi$-axions participate in the parity-even portal.  It was observed in \cite{Arvanitaki:2014faa} that a simple $F^2$ coupling induces a small oscillation in the fine structure constant $\alpha_e$. Here we apply this to the $\pi$-axion scenario.

Consider homogeneous field profiles similar to the standard axion case, 
\begin{equation}\label{classicalfieldprofiles}
\begin{split}
    \pi^r_i(t)&=\pi^r_{i,0}\cos(m_it+\delta_i), \\
    \pi^c_j(t)&=\pi^c_{j,0}e^{i\theta_j}\cos(m_jt+\delta_j),
\end{split}
\end{equation}
where the $r$ and $c$ superscripts denote real or complex, the $\pi^{r/c}_0$ are dimensionful constants which we take to be real, the $\theta_j$ are global phases, and the $\delta_j$ are initial, random phases.

The interactions 
formally include a sum over species but for simplicity, let us discuss the diagonal terms $\sim |\pi|^2F^2$. The correction to the fine structure constant is \cite{VanTilburg:2015oza} 
\begin{equation}\label{alphae}
    \alpha_e(t)=\alpha_e\Big(1+\frac{2\lambda e^2}{\Lambda^2}\varepsilon^2 \sum_i|\pi_{i,0}|^2\cos^2(m_it+\delta_i)\Big),
\end{equation}
for each species that still is in abundance today (the factor of $\lambda/\Lambda^2$ in Eq.~\eqref{alphae} is $\lambda_3/\Lambda_3^2$ for charged species and $\lambda_4/\Lambda_4^2$ for neutral species).
The amplitudes are related to the local dark matter density as 
\begin{equation}\label{amp}
    \pi_{i,0}\simeq \frac{\sqrt{2\rho^i_{\text{DM}}}}{m_{\pi_i}},
\end{equation}
where $\rho^i_{\text{DM}}$ is the contribution from $\pi_i$. As is clear from Fig.~\ref{fig:spectra}, the lightest $\pi$-axions will contribute most to the local energy density, thus their amplitudes will be largest contribution to Eq.~\eqref{alphae} according to
Eq.~\eqref{amp}. 

The experiment discussed in \cite{VanTilburg:2015oza} states that the oscillations in $\alpha_e$ are potentially observable for an amplitude as small as $10^{-16}$, which translates to the requirement
\begin{equation}
    \frac{\rho^i_{\text{DM}} \varepsilon^2 }{\Lambda^2m_{\pi_i}^2}\gtrsim 10^{-15}.
\end{equation}
A recent estimate for the local dark matter density is given 
in \cite{Sofue:2020rnl} as $\rho_{\text{DM,loc}}\sim 0.39\pm0.09$ GeV/cm$^3$. For the example of `fuzzy' dark matter, with $m=10^{-21} $eV, composing all of the relic density, then $\Lambda 
< 10^8 \varepsilon $ GeV in order to observable. Assuming $\Lambda \sim m_{W} \sim g v$,  with $v\sim F_{\pi} \sim 10^{16}$ GeV for fuzzy dark matter, this requires a small gauge coupling $g< 10^{-8}$. There is a broad parameter space of $\pi$-axion mass, millicharge $\varepsilon$, and dark W boson mass $m_{W}$, where the oscillations can be in the observable range.

Our model is distinct from standard axion case again due to the potential for multiple contributions in the sum Eq.~\eqref{alphae}, which encodes the tightly packed discretum of $\pi$-axion masses. Any observed oscillations in $\alpha_e$ would not be able to be fit with a single frequency or amplitude, even in the case where the dark quarks are fully degenerate, as the initial phases $\delta_i$ as well as the  masses, will be different.

\subsection{Parametric Resonance of Photons in Axion Stars}

Finally, we consider the impact of the coherent oscillations of the $\pi$-axion fields on excitations of the Standard Model photon field. As discussed in \cite{Marsh:2015xka}, the axion couplings to photon can lead to a parametric resonance of the latter, which is enhanced in dense environments. This phenomenon is well studied in the case of conventional axions, see Refs.~\cite{Visinelli:2017ooc,Amin:2020vja,Amin:2021tnq,Du:2023jxh,Escudero:2023vgv,Chung-Jukko:2023cow}, where it can allow for indirect detection via various astrophysical signatures.
Parametric resonance has also been studied in the context of ultralight millicharged dark matter in \cite{Jaeckel:2021xyo}. Additionally, the gravitationally-induced resonance of axion fields and gravitons has been studied in  \cite{Alsarraj:2021yve,Brandenberger:2023idg}.  Here we study the possibility of parameter resonance production of photons in the $\pi$-axiverse.

Both parity-odd and parity-even portals will contribute to parametric resonance of photons.  Including the sum over $\pi$-axion species, the Lagrangian including all interactions for the SM photon is 
\begin{equation}\label{LU(1)}
\begin{split} 
   & \mathcal{L}^{\mathrm{int}}_{U(1)_\text{EM}}= \\
   &-\varepsilon^2\Big(\frac{\lambda_3}{2\Lambda_3^2}\sum_{i,j}^{\text{charged}}\pi_i \pi_j^*+\frac{\lambda_4}{2\Lambda_4^2}\sum_{i,j}^{\text{neutral}}\pi_i\pi_j^*\Big)F_{\mu\nu}F^{\mu\nu} \\
    & -\frac{\lambda_1}{2F_\pi}\varepsilon^2\sum_i^{\text{neutral,}\mathbb{R}}\pi_iF_{\mu\nu}\tilde{F}^{\mu\nu}-\frac{\lambda_2\varepsilon^2e^2}{2}\sum_{i,j}^{\text{charged}}\pi_i\pi_j^*A_\mu A^\mu \\
    &+\text{h.c.}
    \end{split}
\end{equation}
We then take the Fourier representation of $\vec{A}$ as
\begin{equation}
    \vec{A}(t,\vec{x})=\sum_{\pm}\int\frac{d^3k}{(2\pi)^3}e^{i\vec{k}\cdot\vec{x}}\hat{\varepsilon}_{\vec{k},\pm}A_{\vec{k}\pm}(t)+\text{c.c.},
\end{equation}
and for the $\pi$-axions, we assume the field profiles Eq.~\eqref{classicalfieldprofiles}. The equation for the mode functions is then of the form

\begin{equation}\label{modeEOM}
\begin{split} 
0&=    \Big(1+P(t)\Big)\Big(A''_{\pm}+k^2A_{\pm}\Big)+B(t)A'_{\pm} \\
&+\Big(C_\pm(t)k+D(t)\Big)A_{\pm},
\end{split} 
\end{equation}
where explicitly,

\begin{equation}\label{Pcoeff}
    \begin{split}
         P(t)&=\frac{4\lambda_3}{\Lambda_3^2}\varepsilon^2\sum_{i,j}^9\pi^c_{i,0}\pi^c_{j,0}\cos(\theta_i-\theta_j) \cos\varphi_i(t)\cos\varphi_j(t)\\
        &+\frac{2\lambda_4}{\Lambda_4^2}\varepsilon^2\Big[2\sum_{i,j}^{6}\pi^c_{i,0}\pi^c_{j,0}\cos(\theta_i-\theta_j) +\sum_{i,j}^5\pi^r_{i,0}\pi^r_{j,0}\\
        &+4\sum_{i=1}^5\sum_{j=1}^6\pi_{i,0}^r\pi_{j,0}^c\cos\theta_j\Big] \cos\varphi_i(t)\cos\varphi_j(t)
    \end{split}
\end{equation}
\begin{equation}\label{Bcoeff}
    \begin{split}
        B(t)&=-\frac{4\lambda_3}{\Lambda_3^2}\varepsilon^2\sum_{i,j}^9\pi^c_{i,0}\pi^c_{j,0}\cos(\theta_i-\theta_j) \\
        &\times \Big(m_i\sin\varphi_i(t)\cos\varphi_j(t)+m_j\cos\varphi_i(t)\sin\varphi_j(t)\Big) \\
        &-\frac{2\lambda_4}{\Lambda_4^2}\varepsilon^2\Big[2\sum_{i,j}^{6}\pi^c_{i,0}\pi^c_{j,0}\cos(\theta_i-\theta_j) +\sum_{i,j}^5\pi^r_{i,0}\pi^r_{j,0}\\
        &+4\sum_{i=1}^5\sum_{j=1}^6\pi_{i,0}^r\pi_{j,0}^c\cos\theta_j\Big]  \\
        &\times \Big(m_i\sin\varphi_i(t)\cos\varphi_j(t)+m_j\cos\varphi_i(t)\sin\varphi_j(t)\Big),
        \end{split}  
\end{equation} 
along with
\begin{equation}\label{Ccoeff}
        C_{\pm}(t)=\pm\frac{2\lambda_1}{F_\pi}\varepsilon^2\sum_i^5\pi^r_{i,0}m_i\sin\varphi_i(t),
\end{equation}
defining the shorthand $\varphi_i(t)=m_it+\delta_i$. In both (\ref{Pcoeff}) and (\ref{Bcoeff}), the sum over the first line is the nine charged species, the other three sums are over the six complex and five real neutral species, while the sum in (\ref{Ccoeff}) is over the five real, neutral species. The $D(t)$ function is the sum over charged species,
\begin{equation}\label{Dcoef}
    D(t)=2\lambda_2\varepsilon^2e^2\sum_{i,j}^9\pi^c_{i,0}\pi^c_{j,0}\cos(\theta_i-\theta_j)\cos\varphi_i(t)\cos\varphi_j(t).
\end{equation}

 Inserting these functions into Eq.~\eqref{modeEOM} results in a special form of Hill's equation, more complicated than the Mathieu equation generally encountered in axion cosmology.

In the most general case where all of the $\pi$-axion masses differ, there will clearly be an exceedingly rich resonant structure due to the different in driving frequencies and amplitudes. However, even in the case where the dark quark masses are fully degenerate, the parametric resonance will differ from the standard axion scenario \cite{Arza:2020eik,Sigl:2019pmj,Tkachev:1987cd,Tkachev:2014dpa,Arza:2018dcy,Hertzberg:2018zte}. The sums in Eqs.~\eqref{Bcoeff} and (\ref{Ccoeff}) slightly simplify due to Eq.~\eqref{amp}; following the discussion of mass hierarchies above, there are just three mass scales for the neutral $\pi$-axions. Since the interactions turn off when $\varepsilon=0,$ we generically have 
nine unique masses for the charged $\pi$-axions. In principle, the numerator in Eq.~\eqref{amp} can be different for all stable $\pi$-axions, even if they have the same masses, and we also have no reason to expect the initial phases $\delta_i$ to be related. In summary, even in the simplest case where the quark masses are fully degenerate, 
the model predicts a very rich resonant structure. 

In scenarios where the frequency of the produced photons is observable, as in \cite{Amin:2020vja}, the $\pi$-axiverse predicts a rich signal due to this resonance structure. The combined effect of these resonances would also accelerate the instability of axion stars. 
We leave these possibilities to future work.

\section{Conclusion}
\label{sec:discussion}

In this work we have developed the theory, cosmology, and observables of the $\pi$-axiverse. The $\pi$-axiverse joins the string axiverse as a possible origin for ultra-light axion-like particles, which can be probed by a variety of cosmological and direct approaches, see \cite{Marsh:2015xka} for a review. In its simplest form, the $\pi$-axiverse emerges from a dark Standard Model wherein all 6 flavors of quark are ultralight. For $N_f$ flavors of light quark, this leads to $N_f ^2 - 1$ ultralight $\pi$-axions. This scenario leads to multiple species of axion dark matter characterized by a spectrum masses and a  mixture of parity-odd and parity-even couplings, leading to complementary signals in a range of observable windows. The parameter space for this class of models is large and the signatures may be diverse. The authors are currently working on a follow-up examination of that parameter space the production of further figures and discussion to guide study of the model.

This scenario offers many interesting directions for future work:\\

{\it Fuzzy $\pi$-axion Dark Matter:} An ultralight boson with mass $m\sim 10^{-21}$, known as Fuzzy dark matter \cite{Hu:2000ke}, is a compelling resolution of the core-cusp problem \cite{Bullock:2017xww,Burkert:1995yz}. However, this scenario faces tight constraints, in particular from stellar heating due to wave interference \cite{Dalal:2022rmp,Amin:2022pzv}. The latter can be ameliorated by introducing a large number of species of fuzzy dark matter fields \cite{Gosenca:2023yjc,Jain:2023ojg}. The $\pi$-axiverse can naturally realize this latter scenario, via degenerate quark masses and vanishing millicharge $\varepsilon$. For an analysis of a scenario with three $\pi$-axions, we refer the reader to \cite{Maleknejad:2022gyf}.\\

{\it $\pi$-axion Self-Interactions:} A distinguishing feature of the $\pi$-axiverse are the interactions not only with the Standard Model, but also self-interactions of $\pi$-axions, e.g. quartic interactions $\pi^4$ and $|\pi_i|^2 |\pi_j|^2$. Self-interactions of ultralight dark matter can play an important role in structure formation \cite{Mocz:2023adf}, boson stars \cite{Croon:2018ybs}, and galactic dynamics \cite{Glennon:2022huu,Glennon:2020dxs}. The $\pi$-axion self-interactions from dark QCD physics are given by the higher-order terms of chiral perturbation theory \cite{Kaplan:2005es,Scherer:2005ri}. The self-interactions receive additional contributions from the dark Standard Model, e.g., after integrating out the dark W bosons. It follows that there is considerable freedom in setting the $\pi$-axion self-interactions, and these could be much larger than the conventional axion ($\sim m^2/f_a ^2$). In addition, the axion quality problem that is thought to plague the QCD axion, wherein the axion mass receives large corrections from higher-order operators, could also be in play here. A future work could consider the approach of \cite{Contino:2021ayn, Podo:2022gyj} where suitable charge assignments can cancel the effects of dangerous operators.\\

{\it Direct Detection of  Charged $\pi$-axions \& Charged Leptons:} For $\varepsilon$ in an intermediate range within $10^{-20} \ll \varepsilon \ll 
 1 $, the charged $\pi$-axions may have a small enough mass yet strong enough coupling to accessible to a variety of experimental probes of millicharged particles, see Tab. I of \cite{Maleknejad:2022gyf}. Similarly, for smaller charged lepton masses, $m\ll F_{\pi}$, the charged lepton masses may be accessible to experiment directly.\\

{\it Direct Detection of Dark Photons:} Throughout this work, we have restricted to the case of $m_{\gamma'}>F_\pi$. However, there is an active experimental effort searching for dark photons at much lower mass scales, see \cite{Caputo:2021eaa} for a review. The detection or constraints on $\epsilon$ impact the model prediction for the charged $\pi$-axion mass, and thus provide a complementary view into the model. \\

{\it Indirect Detection of (Superheavy) Baryons}: As emphasized in \cite{Maleknejad:2022gyf}, the baryons of the dark QCD in this model are a candidate for a superheavy dark matter component: the lightest baryon has mass $m_{b}\sim \Lambda_{\rm dQCD} \sim F_{\pi} \gtrsim 10^{11}$ GeV. 
Superheavy dark matter is characterized by its own array of observables \cite{Carney:2022gse}. By independently probing the  decay constant and millicharge, this provides a complementary window into the $\pi$-axiverse.\\

{\it Embedding within string theory:} Finally, while the $\pi$-axiverse makes no recourse to string theory, we note that it can be straightforwardly realized in variants of the intersecting brane constructions of the Standard Model in Type IIB string theory \cite{Blumenhagen:2005mu}. The dark quarks in this context correspond to open strings connecting two stacks of branes, with identically massless quarks occurring at the intersection. {\it Nearly}-massless quarks, as considered here, therefore arise for nearly-intersecting branes.

This serves a useful purpose of realizing ultralight axion-like particles in the face of the St\"uckelberg mechanism, realized by D3-branes at singularities \cite{Cicoli:2012vw,Cicoli:2013mpa,Cicoli:2013cha,Allahverdi:2014ppa} (for a recent review see Ref.~\cite{Cicoli:2023qri}), wherein an anomalous $U(1)$  `eats' the axion to become massive through the St\"uckelberg mechanism.\\

We leave these and other interesting directions to future work.

\begin{center}
{\bf ACKNOWLEDGEMENTS }
\end{center}
 The authors thank Mudit Jain, Matthew Johnson, David Kaiser, Eiichiro Komatsu, Andrew Long, David Marsh, Guillaume Payeur, Wenzer Qin, and Michael Toomey for helpful comments. SA and TM are supported by the Simons Foundation, Award 896696. HBG is supported by the National Science Foundation, MPS Ascending Fellowship, Award 2213126. EM is supported in part by a Discovery Grant from the Natural Sciences and Engineering Research Council of Canada.

 \bibliographystyle{JHEP}
\bibliography{refs.bib}

\end{document}